\documentclass[10pt]{article}
\usepackage{epsfig}
\usepackage{a4,isolatin1,amssymb,amsmath}
\begin{document}
\centerline{\Large Ensemble versus Individual System in Quantum
        Optics\footnote{Invited lecture at Fundamental Problems in Quantum
  Theory Workshop, August 3-7, 1997, University of Maryland Baltimore
  County}}
\begin{center}
{\em Gerhard C. Hegerfeldt\cite{email}}\\
Institut für Theoretische Physik,
Universit{\" a}t G{\" o}ttingen, Germany
\end{center}
\begin{abstract}
Modern techniques allow experiments on a single atom or system, with
new phenomena and  new challenges for  the theoretician.  
We discuss what quantum mechanics has to say about a single system.
The quantum jump approach as well as the role of
quantum trajectories are outlined and a rather sophisticated example
is given.
\end{abstract}
\vspace*{1cm}
\noindent {\large \bf 1. Introduction}
\vspace*{0.7cm}

Until a decade or so ago only experiments involving {\em many} atoms
were possible, e.g. atoms in an atomic beam or in a gas. With a beam one
would have a repetition of measurements on an ensemble, while experiments
on atoms in a gas --- dilute and with no cooperative effects --- can
often be viewed as a simu  ltaneous measurement on an ensemble.

This is well adapted to the statistical interpretation of quantum
mechanics. For the purpose of the present lecture 
a quantum mechanical expectation value is understood as a {\em mean
value} of measurements on systems of an ensemble, i.e. ensemble
averages of an observable or mean square deviations and so on.

With the advent of atom traps, in particular the Paul trap, and with
laser cooling it became possible to store a single atom (ion) --- or
two, three or more --- in a trap for hours or days and to experiment
with it, e.g. study its interaction with light, microwave
radiation or with other atoms. For a single system the statistical interpretation of
quantum mechanics, based on ensembles, is not so readily applicable as
in the case of beam or a gas. The question we want to address here is
the following.
\begin{center}
\framebox[10.4cm][r]{``Does quantum mechanics allow statements for a single
  system ?''}
\end{center}
The answer will be: ``Yes, to some extent.''

Of course this is trivial if the probability in question is 0 or 1. As
a more interesting example of what can happen  consider the
macroscopic dark periods (`electron shelving') of a {\em single}
three-level atom as proposed by Dehmelt \cite{Deh}. The atom is
supposed to have a ground state 1 and two excited states 2 and $2'$
where the former is strongly coupled to 1 and decays rapidly, while
the latter is metastable. The $1-2$ transition is strongly
driven by a laser, and the $1-2'$ transition is weakly driven.

{\em Semiclassically} the behavior of such a single atom is easy to
understand. The electron makes rapid transitions between levels 1 and
2, accompanied by a stream of spontaneous photon emissions, in the
order of $10^8s^{-1}$. These can be detected (and even seen by the
eye; the stimulated emissions are in the direction of the laser). From
time to time the weak driving of the $1-2'$ transition manages to put
the electron into the metastable level $2'$ where it stays for some time
(`shelving'). During this time the stream of spontaneous photons is
interrupted and there is a dark period. Then the electron jumps back to
level 1 and a new light period begins. In an ensemble of
such atoms (e.g. gas with no cooperative effects) light and dark
periods from different atoms will overlap, and consequently one will
just see diminished fluorescence. Only light and dark periods from a
single or a few atoms are directly observable. 

In quantum mechanics, however, the atom will always be in a
superposition of the three state $|1\rangle$, $|2\rangle$, and $|2^\prime\rangle$ 
and never strictly in state $|2^\prime\rangle$ for an extended period,
i.e. there will always be a small admixture of $|2 \rangle$. Since 
$|2\rangle$ decays rapidly the question arises if the dark periods still
occur. Experiments \cite{Darkexp} and early theoretical treatment
\cite{Darktheor} have answered this affirmatively. The duration of the
dark periods is random and can be seconds or even minutes.

To treat problems like these involving a single system Wilser and 
the author \cite{HeWi,Wi,He} developed the quantum jump approach which is
equivalent and simultaneous to the Monte-Carlo wave function approach (MCWF)
\cite{MC}  and to the quantum trajectory approach \cite{QT}. Our
approach is  based on
standard quantum mechanics and nothing new to the latter is added or
required. In the next section we will give a short exposition of the
quantum jump approach with its associated random (`quantum')
trajectories. In Section 3 we discuss the notion of a spectrum in a
light period and of conditional 
spectra which can be defined for a single fluorescing system. In this
case the quantum jump approach leads to more general quantum
trajectories.\\[1cm]
%\vspace{1cm}
{\large \bf 2. The quantum jump approach. Quantum trajectories}
\vspace*{0.7cm}

This approach is based on standard quantum mechanics and adds no new
assumptions or properties to the latter. In many cases it is just a
practical tool for questions concerning a single system and often has
technical and conceptual advantages. More details can be found in
Refs. \cite{HeWi,He,HePle1,HePle2,HePle3,HeSo1} and in the recent
survey \cite{PleKni}.

The underlying idea is that it should make no difference physically
whether or not the photons emitted by an atom are detected and
absorbed once they are sufficiently far away from the atom. 
It therefore suggests itself to employ gedanken photon
measurements, over all space and with ideal detectors, at instances
a time $\Delta t$ apart\cite{Rei}. For a single driven atom this may look as
in Fig. 1. Starting in some initial state with no photons (the laser
field is considered as classical), at  first  one will
detect no emitted photon in space and then at the $n_1$-th measurement
a photon will be detected (and absorbed), the next photon at the
$n_2$-th measurement and so on.\\[.2cm]
%\vspace{0.2cm}
{\em Limits on $\Delta t$:}
\begin{itemize}
\item Ideally one would like to let $\Delta t \to 0$ to simulate
  continuous measurements. But this is impossible in the framework of
  standard quantum mechanics with ideal measurements due to the
  quantum Zeno effect \cite{Zeno}.
\item Intuitively, $\Delta t$ should be large enough to allow
  the photons to get away from the atom.
\item $\Delta t$ should be short compared to level life-times.
\end{itemize}
This leads to the requirement \cite{Dirac}
\begin{center}
\framebox[3.5cm][r]{$\Delta t \cong 10^{-13}-10^{-10}s$} 
\end{center}

To treat these gedanken measurements on a single atom 
we translate them first into an ensemble description as
follows. We consider an ensemble, ${\cal E}$, of many atoms, each with its
own quantized radiation field, of which our individual atom plus field
is a member. At time $t_0=0$ the ensemble is described by the state
$|0_{ph}\rangle|\psi_A\rangle$. Now we imagine that on each member of 
${\cal E}$ photon
measurements are performed at times $\Delta t,..., n \,\Delta
t,...$ . We consider various sub-ensembles of ${\cal E}$:
\begin{eqnarray*}
{\cal E}_0^{(\Delta t)}~ &\equiv& \mbox{ all systems of ${\cal E}$ for
  which at time $\Delta t$ a  photon was detected}\\
{\cal E}_0^{(n\Delta t)} &\equiv& \mbox{ all systems of ${\cal E}$ for
  which at the times $\Delta t,...,~n\Delta t$ no photon}\\
&&\mbox{ was detected (i.e. {\em until} time $n\Delta t$ no 
photon!)}
\end{eqnarray*}
This is depicted in Fig. 2 where our individual system, atom plus
radiation field, is denoted by a dot ${\bf ~\cdot}$ ,
%$\bullet$ \,
 and it is a member of
${\cal E}_0^{(n\Delta t)}$ for $n<n_1$.

Now one can proceed by ordinary quantum mechanics and the von
Neumann-Lüders reduction rule \cite{Lue}. Let ${\cal P}_0$ be the
projector onto the no-photon subspace,
\begin{equation} \label{P}
{\cal P}_0 \equiv |0_{ph}\rangle  {\bf 1}_A  \langle 0_{ph}|
\end{equation}
and let $U(t,t_0)$ be the complete time-development operator,
including the laser driving and the interaction of the atom with the
quantized radiation field. Then the sub-ensemble ${\cal E}_0^{(\Delta t)}$
is described by
\begin{equation} 
{\cal P}_0  U(\Delta t,0)|0_{ph}\rangle |\psi_A\rangle
\end{equation}
 and the sub-ensemble ${\cal E}_0^{(n\Delta t)}$ by
\begin{equation} \label{t}
{\cal P}_0\,U(n\Delta t,(n-1)\Delta t){\cal P}_0...{\cal P}_0\,U(\Delta
t,0)|0_{ph}\rangle|\psi_A\rangle 
\equiv |0_{ph}\rangle |\psi_A\underbrace{(n\Delta t)}_{\equiv t} \rangle~.
\end{equation}
%We will use a coarse-grained time scale to calculate $|\psi_A(t)\rangle$
%further below.

The relative size of the sub-ensemble ${\cal E}_0^{(n\Delta t)}$ is the
probability to find a member of ${\cal E}$ in ${\cal E}_0^{(n\Delta t)}$ and
is given by the norm-squared of the above expression. Hence
\begin{eqnarray}
 P_0(t) & \equiv & \| \,|\psi_A(t)\rangle \|^2   \nonumber \\
& = & \mbox{probability to find no photon until}\,\, t=n\Delta t
\end{eqnarray}
To calculate $|\psi_A(t)\rangle$ we note that
\begin{equation}
{\cal P}_0 \,U(t'+\Delta t,t'){\cal P}_0 = |0_{ph}\rangle
\langle0_{ph}|U(t'
+\Delta t,t')\,|0_{ph} \rangle\langle 0_{ph}|
\end{equation}
and that the inner expression is a purely atomic operator which is
easily obtained by second order perturbation theory. For $\Delta t$
in the above limits  one then obtains, on a coarse-grained time
scale (for which $\Delta t$ is very small), that the time-development
of $|\psi_A(t)\rangle$ is given by a `conditional`, or `reduced`,
non-Hermitian Hamiltonian $H_{\rm cond}$ in the atomic Hilbert space
where, for an $N$-level atom,
\begin{equation}
H_{\rm cond} = H_A(t) - i \hbar \Gamma
\end{equation}
with $ H_A(t)$ the atomic part of the Hamiltonian, including the laser
driving, and
\begin{eqnarray}
\Gamma &\equiv& \sum_{\alpha ij \atop \alpha <i,j}
\Gamma _{i \alpha \alpha j}
|i \rangle \langle j|\\
\Gamma_{ijlm} &\equiv&
e^2 \langle i|{\bf X}|j \rangle \cdot \langle |{\bf X}|m \rangle
 \,|\omega_{lm} |^3/(6\pi \epsilon_0 \hbar c^3)
\end{eqnarray}
and $e{\bf X}$ the dipole operator. $\Gamma$ consists of generalized damping
terms, and we note that
$$
\Gamma _{i \alpha \alpha i}\equiv \Gamma _{i \alpha}\ = \frac{1}{2} A_{i\alpha}
$$
where $A_{i \alpha}$ is the Einstein coefficient for the transition 
from level $i$ to level $\alpha$. Thus, on a
coarse-grained time scale, one obtains
\begin{eqnarray}
|\psi_A(t)\rangle & = &
{\cal T} \exp\{ - i\int_{0}^{t}dt'H_{\rm cond}(t')/\hbar \}
|\psi_A (0)\rangle \\
&\equiv& U_{\rm \rm cond}(t,0)|\psi_A(0) \rangle~.
\end{eqnarray}
In an obvious extension to density matrices,
\begin{equation}
\rho_A^0(t) \equiv U_{\rm cond}(t,0) \rho_A(0) U_{\rm cond}(t,0)^\dagger
\end{equation}
describes the sub-ensemble 
with no photon detection until time $t$, with the corresponding no-photon 
probability given by
\begin{equation}
P_0(t) = {\rm tr} \rho_A^0(t)~.
\end{equation}

If one lets $\Delta t \to 0$ in Eq. (\ref{t}), with $t = n \Delta
t$ kept fixed, then one easily sees by the same calculation that the
probability to find no photon until time $t$ goes to 1 and that one always
stays in the no-photon subspace. This means that for $\Delta t \to
0$ the dynamics is frozen to the atomic subspace, and this is a
particular form of the quantum Zeno effect.

\vspace*{0.5cm}
\centerline
{\bf A single fluorescent atom as a sample path: Quantum  trajectories}
\vspace*{0.3cm}

Now we can return to the gedanken measurements on our single atom
driven by lasers. We can distinguish different steps in its temporal 
behavior.
\begin{itemize}
\item  Until the detection of the first photon, our atom belongs to
  the sub-ensembles ${\cal E}_0^{(n \Delta t)}$ and hence is
  described by the (non-normalized) vector
\begin{equation}
|\psi_A(t)\rangle = U_{\rm cond}(t,0)|\psi_A(0) \rangle.
\end{equation}
\item The first photon is detected at some (random) time $t_1$,
  according to the probability density
\begin{equation}
w_1(t)= 
- \frac{dP_0(t)}{dt}= - \frac{d}{dt} \|\, |\psi_A(t) \rangle \|^2~.
\end{equation}
\item  {\em Jump:} With the detection of a photon the atom has to
be reset to the appropriate state. For example a two-level atom will
be in its ground state right after a photon detection. The general
reset state has been determined in Refs. \cite{He,HeSo1} and it may
depend on $|\psi_A(t_1)\rangle$ where $t_1$ is the detection time.
\item From this reset state the time development then continues
  with $U_{\rm cond}(t,t_1)$, until the detection of the next photon
  at the (random) time $t_2$. Then one has to reset (jump), and so on.
\end{itemize}
In this way one obtains a
\begin{center}
\framebox[7.8cm][r]{stochastic path in the Hilbert space of the atom.}
\end{center}
The stochasticity of this path is governed by quantum theory, and the
path is called a {\em quantum trajectory}.
The stochastic process underlying these trajectories is a jump process
with values in a Hilbert space. If the reset state is always the same,
e.g. the ground state, one has a renewal process. If the reset state
depends on $|\psi_A(t_i) \rangle$ one has a Markov process only.

As shown in Ref. \cite{He} the ensemble of all possible trajectories
obtained in this way leads to a reduced density matrix for the ensemble of
atoms which satisfies the usual optical Bloch equations. 
This is a nice consistency check \cite{So}. 

In case of a renewal process the parts of a trajectory between jumps
behave like an ensemble created by {\em repetition} from a single
system at stochastic times. In the general case the reset states can
all be different so that this repetitive property is no longer true.

\vspace*{.5cm}
\centerline
{\bf Observables for a single system}
\vspace*{0.3cm}

Since the individual photon detection times for a single driven atom
are random, they cannot be predicted. However, time averages along a
trajectory are more promising, e.g. the mean distance between two
subsequent photon detections or other correlations.

If the underlying stochastic process is {\em ergodic} then 
\begin{center}
\framebox[9cm][r]{time average over a single trajectory = ensemble
  average}
\end{center}
and this equality allows easy calculation. In many cases, e.g. for a
renewal process, ergodicity is easy to see. We believe that it is
 probably true in general for the quantum
trajectories\cite{Surprising}.

Hence for {\em observables such as time averages quantum mechanics
  allows predictions for single systems}. As applications we mention
  macroscopic dark periods \cite{HeWi,HePle1} and quantum counting
  processes \cite{He} where the axioms introduced by Davies and
  Srinivas \cite{DaSri} are not needed.

There is a word of caution, however. For the observable ``frequency
spectrum of fluorescent radiation'' from a single atom the above
trajectories are not (directly) applicable. This has to do with the
time-energy uncertainty relation. If all photon detection times were
known by measurements then the spectrum would be broadened and
deformed. This shows that the above quantum trajectories are not
``realistic'' and should therefore not be over-interpreted. They are
just a useful quantum mechanical tool in certain situations. There is
also a relation with the consistent-histories approach to quantum
mechanics \cite{Grif}.\\[1cm]
%\vspace*{1cm}
{\large \bf 3. A surprising example. More general quantum trajectories}
\vspace*{0.7cm}

Sometimes it is advantageous to carry over the quantum jump approach
to a  situation where one  asks more general questions about the
temporal behavior of a single system subject to observations. To motivate
this we return to the light and dark periods of the Dehmelt system
mentioned in  the Introduction.
The light and dark periods are depicted in Fig. 3.  The semiclassical
considerations of the Introduction suggest that the light periods are mainly
due to transitions between levels 1 and 2. Now the frequency
distribution of light emitted by a laser-driven two-level system is
given by the Mollow spectrum. For weak driving this consists of a
Lorentzian spectrum around the laser frequency $\omega_L$ 
(``incoherent part'') plus
a $\delta$ peak (Rayleigh peak,
``coherent part'') at $\omega_L$. For strong driving the incoherent
part consists of three Lorentzian parts (Mollow triplet).\\

{\em Complete spectrum:} Let us now turn to the frequency 
spectrum emitted by the
three-level Dehmelt system and use some simple arguments to see what
to expect. {\em Classically,} the $\delta$ or Rayleigh peak of the
two-level system would correspond to the emission of an
electromagnetic field with sharp frequency $\omega_L$. For the
three-level case the amplitude of this field is zero during the dark
periods, and hence  classically one would have an amplitude-modulated
signal. From radio engineering one knows that classical amplitude-modulated
signals contain
sidebands. Because the dark periods have random lengths the sidebands
should be  continuous and should 
lead to a partial broadening of the $\delta$ or Rayleigh
peak. Surprisingly, this is exactly what the fully quantum mechanical
calculation predicts \cite{HePle5}, and the result is given in
Fig. 4. On top of the center  of the Mollow spectrum there is
an additional narrow Lorentzian peak, and then the Rayleigh peak. This is
the spectrum from an ensemble of such atoms, or from a single
atom whose light emitted from time zero to infinity is spectrally
analyzed. For this spectrum one can show that there is no difference
between ensemble and single system, due to ergodicity. Fig. 5 shows
the enlarged center of the spectrum for a different set of parameters.

By the classical analogy, more frequent dark periods should mean more
modulation and should thus lead to a wider broadening of the Rayleigh
peak. The quantum mechanical calculation again confirms this
expectation.\\

 {\em Spectrum for single system in a light period:} 
Taking the classical explanation one step 
further would clearly imply:
\begin{center}
\framebox[7.9cm][r] {
%\begin{center}
spectrum in a light period $\neq$ complete
  spectrum
%\end{center}
}
\end{center}
{\em Remarks}
\begin{itemize}
\item If at all, the notion of 
spectrum in a light period can only be meaningful for a {\em
    single} system!
\item The time-frequency uncertainty relation will of course introduce
  a broadening. The light period under consideration should therefore
  be so long that this broadening is negligible.
\item How does one know one is in a light period? By photon
  counting. But this disturbs the spectrum, by the time-energy
  uncertainty relation, as explained in Section 2! So how to measure
  the spectrum without disturbing it?
\end{itemize}

To overcome the last objection to a proper quantum mechanical notion
of the spectrum in a light period, we suggested in Ref. \cite{HePle5} 
 the setup depicted in Fig. 6. A laser-driven
atom emits radiation. In the right half-space an ideal broadband
photo-detector registers all photons  and triggers a
spectrometer (spectral analyzer) in the following way. A light period
in the right half-space is defined as a sequence of photon detections
whose temporal distance in less than some prescribed time $T_0$. A
dark period is a time interval  which is longer than $T_0$ and 
with no detection occurring. 
Now the broadband counter in the right half-space opens the spectral
analyzer in the left half-space at the beginning of a light period and
closes it at the end (after an additional time $T_0$, to be precise). 
All data for light
periods in the right half-space of length less than some prescribed
$T$
are discarded. In this way one obtains a sequence of spectral data for
the left half-space referring to light periods in the right half-space
of length at least $T$. The spectral data for individual light periods
may depend on the actual detection pattern for the photons in the
right half-space and one can average over these. The notion of spectrum
outlined above may be properly called a conditioned or {\em conditional
spectrum}  because one performs a selection of spectral
data based on prescribed conditions. \\

{\em Generalized quantum   trajectories:}                    
To calculate this conditional spectrum in a light period we have
generalized the above quantum jump approach and its quantum
trajectories in a natural way, adapted to the problem at
hand \cite{HePle5}.
 At a
no-photon detection in the right half-space the state is not projected
with the projector ${\cal P}_0$ of Eq. (\ref{P}) onto the global
no-photon subspace, but rather onto those no-photon states belonging
to modes with momenta in the right half-space. The resulting projected
state will contain, in addition to an atomic part, also photon modes
with momenta in the left half-space. This leads to a more complicated
time development between detections. Once a photon has been detected
and absorbed in the right half-space one has to reset the state
(a ``jump''). The jump in general leads to a density matrix involving
the atoms and photons from the left half-space. In this
way one obtains a quantum trajectory consisting of density matrices
where the latter simultaneously describe the atom together with a
subset of modes of the radiation field.

This generalized quantum jump approach has been used to calculate the
conditional spectrum in a light period \cite{HePle5}. As expected the
broadening of the Rayleigh peak disappears. For the same parameters as
in Fig. 5 the dotted curve in Fig. 7 shows the corresponding spectrum
in a light period which is long enough so that the broadening of the
Rayleigh peak due to the time-frequency uncertainty relation is
negligible. Of course, with increasing length such light periods
become very rare. To define a light period we have chosen $T_0=50
\Gamma_{22}^{-1}$, and so classically a light phase can still have
some amplitude modulation. This explains the small bump in the line
center.

The above example is in several respects amazing. First of all, quite
elementary classical arguments about radiation from driven atoms
turn out to be qualitatively correct (to a large extent also
quantitatively so, cf. Ref. \cite{HePle5}), although the quantum
solution is much more complicated. Furthermore, it shows that there
are interesting questions concerning a single system which at first
sight seem contradictory but still allow a quantum theoretical
treatment.\\[1cm]
{\large \bf 4. Conclusions}
\vspace{0.7cm}

The usual statistical interpretation of quantum mechanics uses the
ensemble point of view. In this view a state vector or density matrix
describes not a single system but an ensemble of identically prepared
systems.  Nevertheless, it is
sometimes useful to ascribe a state $|\psi\rangle$ or density matrix $\rho$
also to a single system. By this we mean that the system is prepared
by the same apparatus as that for the corresponding ensemble.

In this lecture we have tried to convey several points.
\begin{itemize}
\item Experiments with a single  system (atom, ion) show phenomena
  which are absent for an ensemble (e.g. in a gas with no cooperative
  effects).

\item A convenient tool for a description of such experiments with a
  single system is often given by the quantum jump approach with its
  (random) quantum trajectories. This approach is based on standard
  quantum mechanics and does not go beyond it.

\item The quantum jump approach is useful for questions related to the
  statistics of photons emitted from a single system. More general
  problems, like the frequency spectrum in a light period, require
  more general trajectories.

\item The set of quantum trajectories for a driven atom give the reduced
density matrix equations (optical Bloch equations) for the ensemble of atoms.

\item Since quantum trajectories are adapted to the particular problem
  under consideration a quantum trajectory is not a ``realistic''
  property of a single system. Rather, the complete information is
  contained in the state vector (or density matrix) of the system plus
  radiation field.
\end{itemize}

An interesting question relating to ergodicity was touched upon in the
text. Are time averages independent of the particular quantum
trajectory? Or could they vary for different individual systems? That
they do not is usually assumed in experiments. Is this always true or
does one need additional assumptions?

\newpage
\centerline{Figure Captions
\\ [1cm]}

Fig. 1: Repeated photon measurements.\\[1cm] 

Fig. 2: Ensemble $\cal E$ and sub-ensembles. The dot denotes our single
system.\\[1cm]

Fig. 3: The light periods consist of rapid sequences of photon
emissions.\\[1cm]

Fig. 4: Frequency spectrum  with an additional narrow Lorentzian 
peak due to dark periods ($\Delta \equiv \omega - \omega_L$).\\[1cm]

Fig. 5: Enlarged center of spectrum for a different set of
parameters ($\Delta \equiv \omega - \omega_L$).\\[1cm]

Fig. 6: The broadband photo-detector in the right half-space triggers
the spectrometer in the left half-space during a long light
period.\\[1cm]

Fig. 7: Dotted line: The narrow Lorentzian peak of Fig. 5 
is absent in a long light period ($\Delta \equiv \omega - \omega_L$).

\begin{thebibliography}{99}
\bibitem{Surprising} It would indeed be very interesting if there were
  non-ergodic quantum trajectories. Then there could be different time averages
  for different trajectories.
\end{thebibliography}
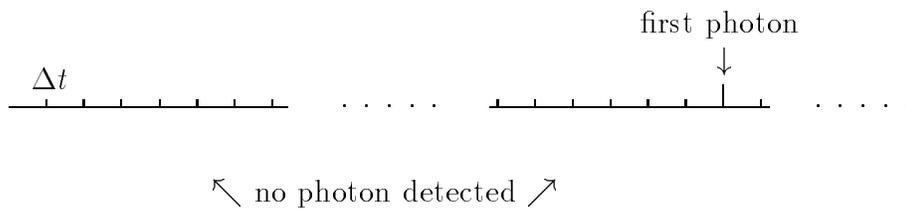
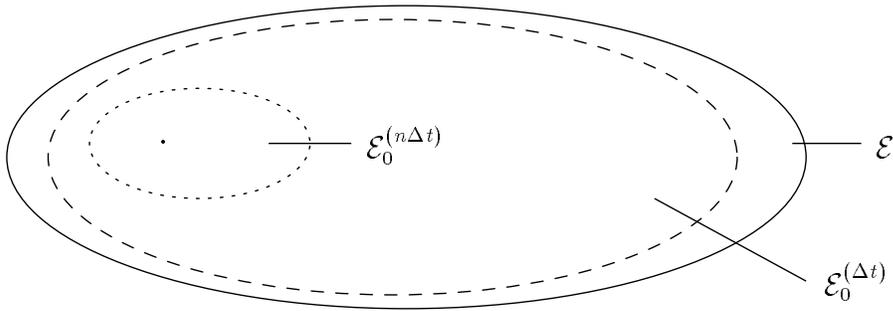
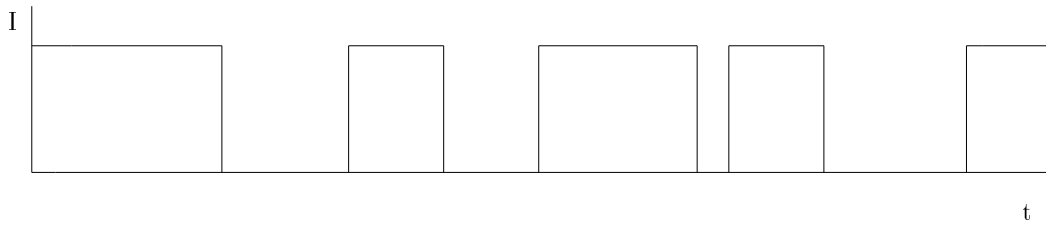
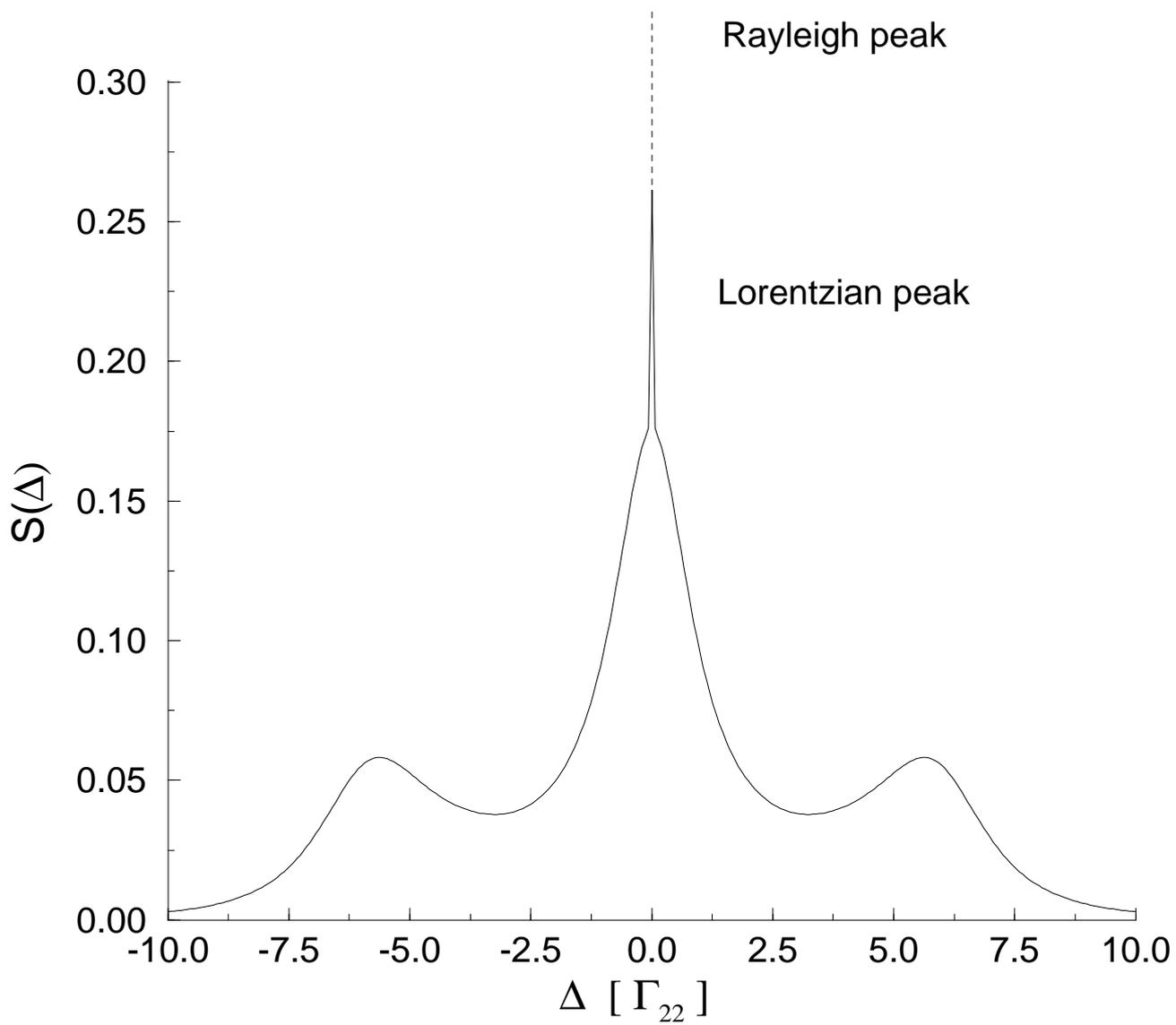
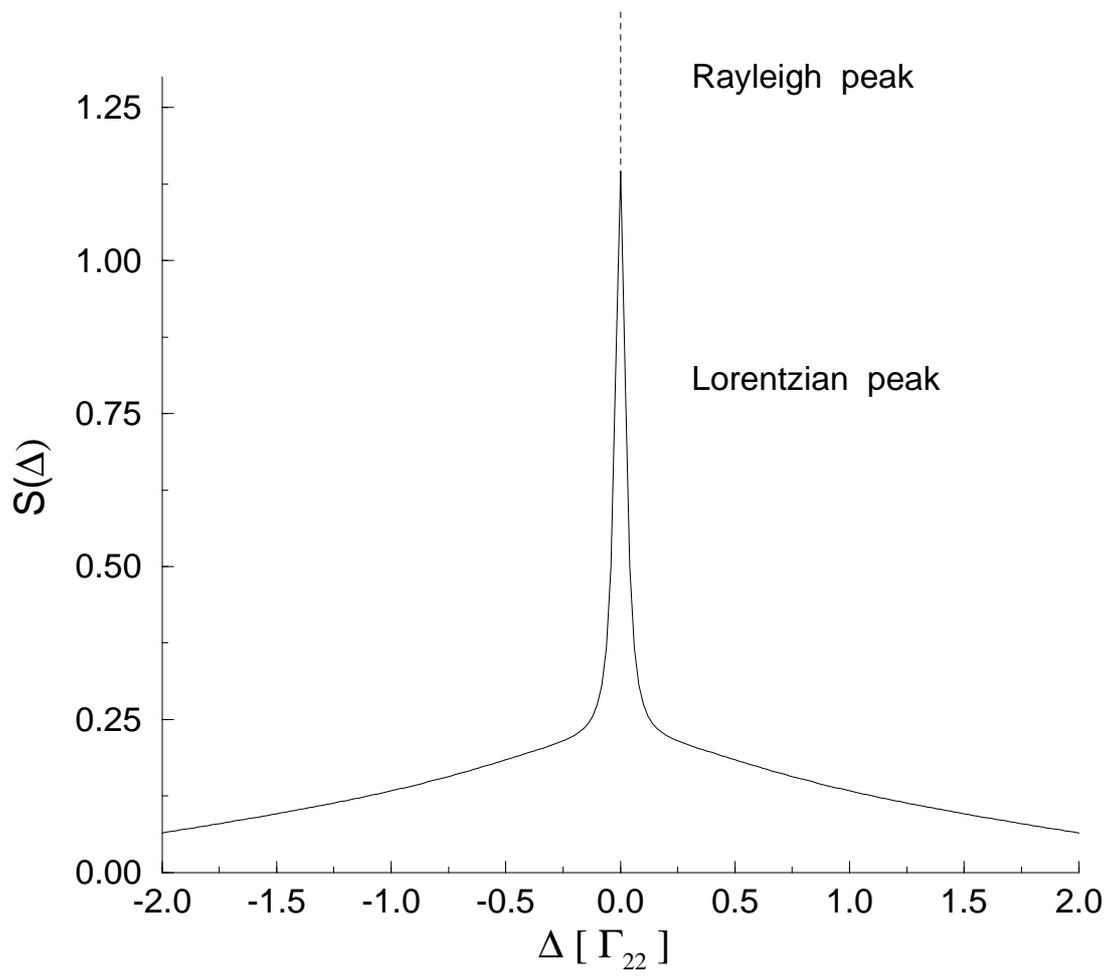
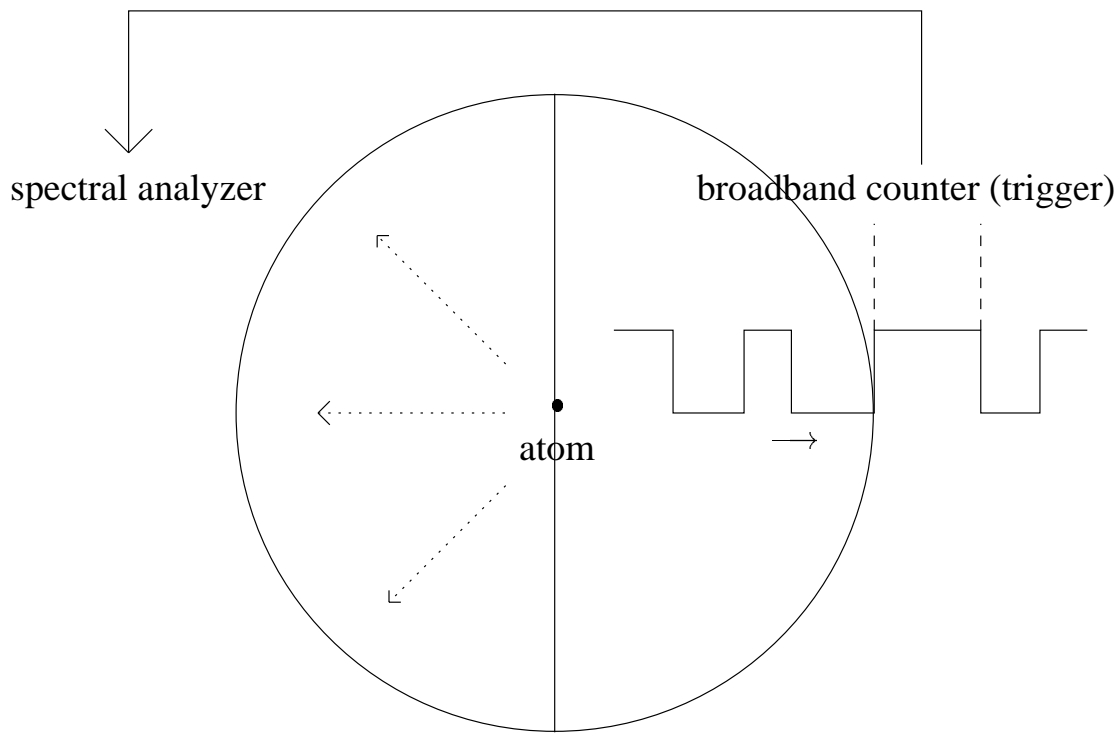
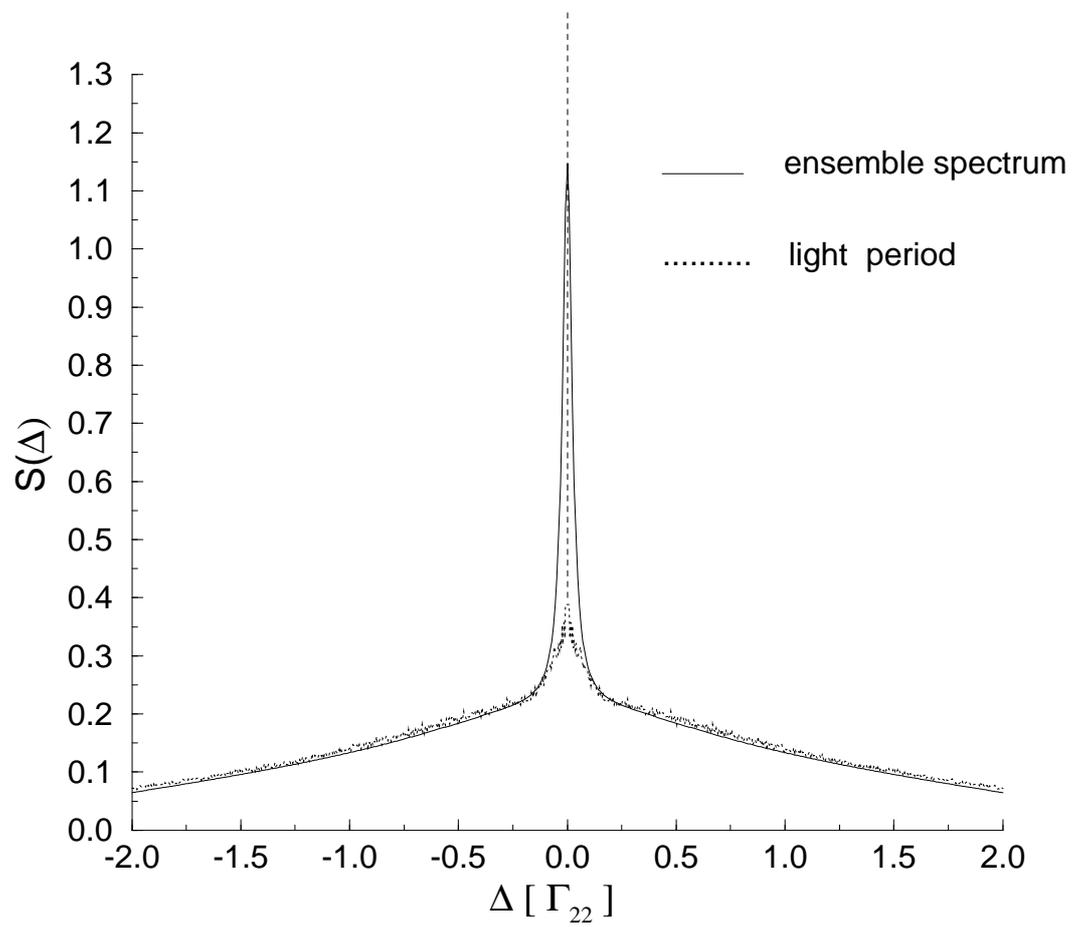
\end{document}